\documentclass[aps,prl,twocolumn,superscriptaddress]{revtex4-2}

\usepackage[colorlinks=true,linkcolor=blue,citecolor=blue,urlcolor=blue]{hyperref}
\usepackage{amsmath}
\usepackage{graphicx}
\usepackage{bm}
\usepackage{comment}
\newcommand{\cwps}{{CuWP$_2$S$_6$}}

\begin{document}

\title{Sign control of photocurrents by spin-group-symmetry breaking \\
in altermagnetic insulators}

\author{Gastón Blatter}
\affiliation{Institute for Theoretical Solid State Physics, IFW Dresden, Helmholtzstrasse 20, 01069 Dresden, Germany}
\author{Xiao Zhang}
\affiliation{Institute for Theoretical Solid State Physics, IFW Dresden, Helmholtzstrasse 20, 01069 Dresden, Germany}
\affiliation{State Key Laboratory of Ultra-intense laser Science and Technology, Shanghai Institute of Optics and Fine Mechanics (SIOM), Chinese Academy of Sciences (CAS), Shanghai 201800, China}
\author{Jeroen van den Brink}
\affiliation{Institute for Theoretical Solid State Physics, IFW Dresden, Helmholtzstrasse 20, 01069 Dresden, Germany}
\affiliation{Würzburg-Dresden Cluster of Excellence ctd.qmat, Germany}
\author{Mengli Hu}
\thanks{m.hu@ifw-dresden.de}
\affiliation{Institute for Theoretical Solid State Physics, IFW Dresden, Helmholtzstrasse 20, 01069 Dresden, Germany}
\author{Shu Zhang}
\thanks{shu.zhang@oist.}
\affiliation{Collective Dynamics and Quantum Transport Unit, Okinawa Institute of Science and Technology Graduate University, Onna-son, Okinawa 904-0412, Japan}
\affiliation{Institute for Theoretical Solid State Physics, IFW Dresden, Helmholtzstrasse 20, 01069 Dresden, Germany}
\newcommand{\s}[1]{\textcolor{magenta}{#1}}
\newcommand{\ML}[1]{\textcolor{blue}{#1}}
\newcommand{\g}[1]{\textcolor{red}{#1}}

\date{\today} 

\begin{abstract}
Controlling physical responses through symmetry breaking is a central paradigm in quantum materials, enabling novel functionalities. Here we determine the effects of spin-group-symmetry breaking on nonlinear optical responses of collinear altermagnetic insulators. 
Using shear strain as an example, we show that the direction of symmetry-breaking induced components of charge and spin photocurrents are locked to the sign of the strain.
In the absence of spin-orbit coupling, this effect is intuitively captured by the spin-gap asymmetry|an imbalance between spin-up and spin-down direct band gaps which couples trilinearly with the Néel order and the strain.
We demonstrate this mechanism with density functional theory calculations on the recently proposed altermagnet CuWP$_2$S$_6$. 
Having symmetry-guided control of both charge and spin photocurrents allows, vice versa, to reveal and investigate altermagnetism in insulating materials by exploration of their optical responses.
\end{abstract}


\maketitle


\textit{Introduction---}Altermagnets are a class of magnetic materials that exhibit spin-split bands with zero net magnetization~\cite{PhysRevX.12.040501, PhysRevX.12.031042}. The combination of large spin currents, ultrafast spin dynamics, and the absence of stray magnetic fields makes them particularly promising candidates for spintronic applications~\cite{https://doi.org/10.1002/adfm.202409327,PhysRevLett.126.127701,Jungwirth2025}. To date, many bulk systems have been experimentally confirmed to be altermagnetic~\cite{Amin2024,Yamamoto2025,PhysRevLett.132.036702,Reimers2024}, while the two-dimensional (2D) limit is under active exploration. Several interesting phenomena have recently been theoretically proposed in 2D altermagnets, including the type-II quantum spin Hall state~\cite{Zhang2025,Wei2025,Chen2025_QSH_Altermagnetic}, electric-field-switchable altermagnets~\cite{PhysRevLett.134.106801,Wang2024}, magnon enhanced superconductivity~\cite{Brekke2023}, suppression of Kondo screening~\cite{Diniz2024}, sliding induced anomalous valley Hall effect~\cite{Li_Zhang2025} and excitonic effects~\cite{Cao2025, Sun2025}.
Experimentally, while spin-polarized currents and Hall-like responses provide established probes of metallic altermagnets~\cite{rv1n-vr4p, Naturecomm2025, Farajollahpour_Ganesh_Samokhin_2025, PhysRevB.111.184407}, the absence of a Fermi surface in insulating altermagnets suppresses conventional transport signatures, leaving open the question how the spin-resolved electronic structure in altermagnetic insulators (AMIs) can be probed and controlled.
These properties make the 2D AMIs particularly interesting for fundamental study.

Nonlinear optical responses (NLORs) provide an approach that remains operative in insulating systems, as they can arise from interband transitions. Second-order optical phenomena such as shift and injection (spin) currents~\cite{bpve1981,Aversa1995,sipe2000,ahn_low-frequency_2020,morimoto2016,dejuan2017, nagaosa2017, konig2017, kim2017, zhang2018, parker2019, Orenstein_2021} are highly sensitive to band geometry and symmetry, serving as a powerful tool in revealing topological properties and symmetry breaking in a wide range of nonmagnetic and magnetic insulators~\cite{Xu2019,Liu2025,Swain2019,Li2025,Canfield2004}. 

For the investigation of NLORs in AMIs, symmetry analysis is fundamental, as it often is the first step in understanding and predicting physical responses in quantum materials. For instance, second-order optical responses are altogether forbidden in centrosymmetric systems and can arise only when inversion symmetry is broken~\cite{sipe2000,ahn_low-frequency_2020}. 
In altermagnets, symmetry constraints acquire an enriched structure through spin-group symmetries, in which spatial operations are composed with spin rotations~\cite{PhysRevX.12.021016,10.21468/SciPostPhys.18.3.109,PhysRevX.14.031037,PhysRevX.14.031038,PhysRevX.14.031039,Shinohara:ib5119,Litvin:a14103, Litvin_Opechowski_1974a}. Physical responses in altermagnets are therefore governed not only by crystal symmetries but also constrained by the interplay between spin and lattice degrees of freedom. 
Given the general correspondence between symmetry breaking and allowed physical responses, it is natural to ask what new phenomena can emerge by breaking spin-group symmetries. In particular, since nonlinear optical responses are highly symmetry selective, they provide an ideal setting to explore new routes for manipulating spin and charge photocurrents in altermagnets through controlled symmetry breaking, while offering further insights into the role of spin-group symmetry in AMIs.


\begin{figure}
    \centering
    \includegraphics[width=\linewidth]{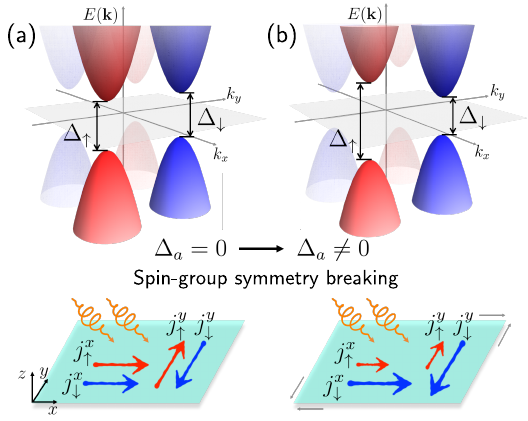}
    \caption{(a) Schematic band structure of a two-dimensional $d$-wave altermagnetic insulator, where red (blue) denotes the spin up (down) bands. The rectified photocurrents obey corresponding symmetry constraints, which, in this example, forbid a spin current along $x$ and charge current along $y$. (b) Upon spin-group symmetry breaking induced by shear strain, the band structure is shifted and develops a finite spin-gap asymmetry $\Delta_a$. The spin and charge photocurrent components forbidden in (a) are are thereby activated.}
    \label{fig:fig1}
\end{figure}


Determining the evolution of NLORs of collinear AMIs under spin-group symmetry breaking, we show here that lattice perturbations such as shear strain activate previously forbidden second-order charge and spin photocurrents, whose direction is deterministically locked by symmetry to the sign of the applied perturbation. The spin-group symmetry breaking features the lifting of the symmetry-enforced equivalence between spin-resolved gaps, which we define as the spin-gap asymmetry (SGA). As we will establish below, this quantity plays a role analogous to magnetization in symmetry analysis of collinear altermagnets and provides the key control parameter governing their NLORs.
We demonstrate this general mechanism by performing first-principles calculations on the antiferroelectric $d$-wave altermagnet, CuWP$_2$S$_6$ in the form of a monolayer~\cite{PhysRevLett.134.106801}, and explicitly show strain-induced sign reversal of nonlinear photocurrents. Our results identify NLORs as a sensitive probe for symmetry breaking in AMIs and establish a symmetry-based route to controllable spin and charge photocurrents in AMIs.

\textit{Main concepts---} 
In an insulator, the absence of a Fermi surface implies that light-induced charge and spin photocurrents originate predominantly from interband excitations between valence and conduction bands.
Our focus here is on insulating collinear altermagnets, where the electronic structure can be decomposed into independent spin channels labeled $\uparrow$ and $\downarrow$. While spin-orbit coupling may be present, its effect on spin splitting and mixing is taken to be small, as consistent with the defining characteristics of altermagnets, and will be neglected.
Let us consider the symmetry constraints imposed on the band structure, which leads to a $d$-wave spin splitting as illustrated in Fig.~\ref{fig:fig1}(a). A direct consequence is the equivalence between the spin-resolved (minimal) direct gaps $\Delta_{\uparrow/\downarrow}$:
\begin{equation}
\begin{split}
\Delta_\uparrow &=E^{\uparrow}_\text{CBM}(\mathbf{k}_0) - E^{\uparrow}_\text{VBM}(\mathbf{k}_0) \\
&\overset{g}{=} E^{\downarrow}_\text{CBM}(g_l\mathbf{k}_0) - E^{\downarrow}_\text{VBM}(g_l\mathbf{k}_0) = \Delta_\downarrow,
\end{split}
\end{equation}
which is enforced by the spin-group symmetry $g=[g_s||g_l]$, where $g_s$ flips the spin and $g_l$ connects the magnetic sublattices. Here, $\mathbf{k}_0$ and $g_l\mathbf{k}_0$ denote the momenta at which the spin-up and spin-down direct gaps occur, respectively. Notably, these two momenta need not coincide in altermagnets, in contrast to PT-symmetric N\'eel antiferromagnets.
Depending on the specific spin-group symmetry class~\cite{PhysRevX.12.040501, PhysRevX.12.031042}, the spin gaps exhibit various momentum-resolved symmetry patterns.

Breaking the spin-group symmetry $g$ gives rise to the spin-gap asymmetry (SGA), which we define as
\begin{equation}
\Delta_a = \Delta_\uparrow - \Delta_\downarrow,
\end{equation}
as illustrated in Fig.~\ref{fig:fig1}(b). 
SGA inherits well-defined symmetry transformation properties: it is invariant under purely spatial operations, and odd (axial) under spin inversion $\Delta_a \overset{[-1||1]}{\longmapsto} \Delta_\downarrow - \Delta_\uparrow  = - \Delta_a$, which implies that $\Delta_a$ transforms according to the product of irreducible representations (IRs) $\Gamma_1 \otimes \Gamma_A^s$ under spin-group symmetry operations. 
Such behavior is analogous to the longitudinal magnetization $M$ (magnetization $\mathbf{M}$ projected onto the direction of the N\'eel order $\mathbf{N}$)~\cite{mcclarty_landau_2024}.

Symmetry-allowed couplings between SGA and external perturbations can be systematically constructed accordingly. In particular, perturbations acting on lattice degrees of freedom can be decomposed based on IRs of the space group. Taking the example of shear strain $\epsilon$, an invariant term in the free energy arises in the form $F\propto \epsilon \Delta_a N$, which is a trilinear coupling between $\epsilon$, the SGA  $\Delta_a$, and the (magnitude of the) N\'eel order $N$. Here, $N$ transforms as $\Gamma_N \otimes \Gamma_A^s$ and $\epsilon$ as $\Gamma_N \otimes \Gamma_1^s$, where $\Gamma_N$ is a one-dimensional nontrivial IR of the group. Such a symmetry-allowed coupling closely parallels the altermagnetic piezomagnetic effect~\cite{mcclarty_landau_2024}. 

The symmetry analysis above establishes SGA as a symmetry-breaking indicator on the band structure level, playing a role analogous to an order parameter in characterizing spin-group symmetry breaking and associated responses in AMIs.
Here, we focus on dc photocurrents arising from second-order optical processes:
\begin{equation}
    j^c = \sigma^{c;ab}(0;\omega,-\omega)E_a(\omega)E_b(-\omega),
    \label{eq:photocurrent_current}
\end{equation}
where $a,b,c$ denotes the spatial directions. In what follows, the current $j^c$ and the optical conductivity tensor $\sigma^{c;ab}$ are also labeled by the polarization of the light (linear/circular), the underlying mechanism (shift/injection)~\cite{ahn_low-frequency_2020, vonbaltz1981, sipe2000, Orenstein_2021} and the spin channel ($\uparrow$/$\downarrow$), and we further define the charge and spin photocurrents as $j^c_\text{charge/spin} = j^c_\uparrow \pm j^c_\downarrow$. The computation formulas of $\sigma^{c;ab}$ are given in Appendix.~\ref{sec:NLORs}. The nonlinear photocurrents are highly sensitive to the spin-group symmetries~\cite{jiang_nonlinear_2025,PhysRevB.111.195210, yjp4-gkj9}. In collinear altermagnets, the symmetry $g_1=[\mathcal{T}U_\textbf{n}(\pi)||E]$~\cite{Khatua2025,PhysRevX.12.021016,schiff2024collinear,10.21468/SciPostPhys.18.3.109} forbids the linear injection Re($\sigma_{\mathrm{inj}}^{c;ab}$) and circular shift Im($\sigma_{\mathrm{shi}}^{c;ab}$) in both spin channels, like in non-magnetic systems~\cite{ahn_low-frequency_2020}. 

\begin{figure*}[t]
    \centering
    \includegraphics[width=\linewidth]{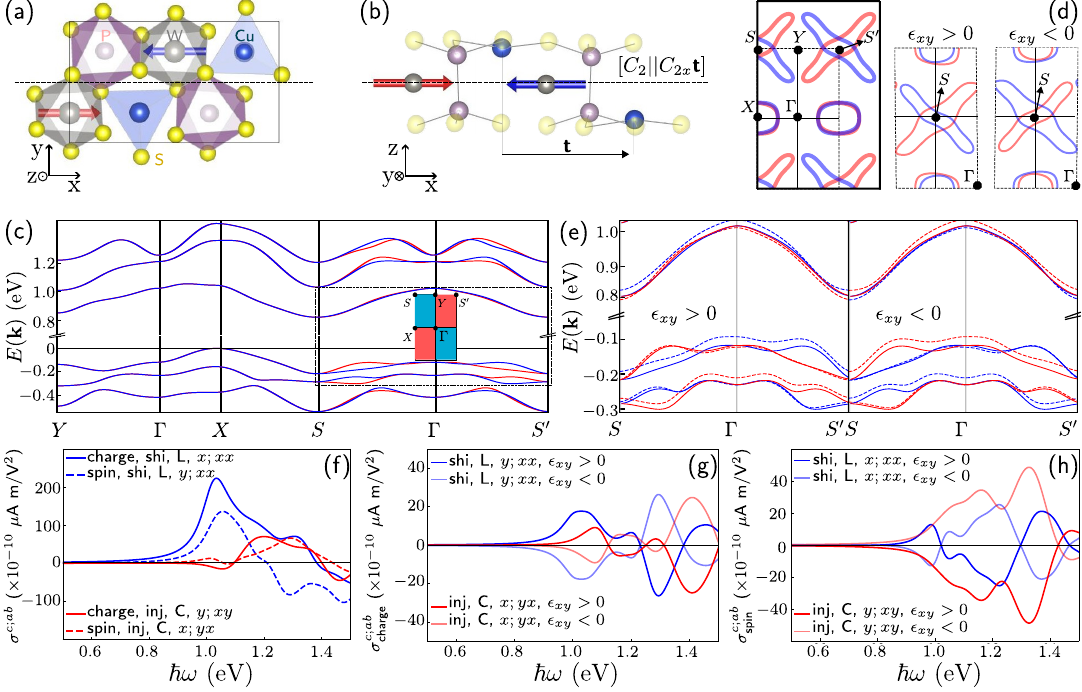}
    \caption{(a–b) Crystal and magnetic structure of the \cwps\ monolayer in the altermanget-antiferroelectric phase.
(c) Band structure of the \cwps\ monolayer without SOC; blue and red lines denote spin-down and spin-up bands, respectively.
(d) Isoenergy contours at $E= E_f-0.2$ eV for pristine and strained monolayers.
(e) Band structures under strain (dashed lines) and without strain (solid lines) plotted for the boxed region in (c).
(f) NLOR of \cwps\ under linearly polarized (blue) and circularly polarized (red) light; solid lines represent charge conductivities, while dashed lines represent spin conductivities.
(g-h) Representative strain-induced charge and spin components in NLOR. The responses are odd upon reversal of shear strain $\epsilon_{xy}$.}
    \label{fig:fig2}
\end{figure*}
Another illustrative example is the $g_2=[C_2\|C_{2x}]$ symmetry in a $d$-wave altermagnet~\cite{chen_unconventional_2025,PhysRevX.14.031038},
which enforces the spin and charge photocurrents to flow along orthogonal directions~\cite{jiang_nonlinear_2025,PhysRevB.111.195210}, following from, e.g.,
\begin{equation}
\sigma_{\uparrow}^{x;xx} \overset{g_2}{\longmapsto} \sigma_{\downarrow}^{x;xx}, \quad
\sigma_{\uparrow}^{y;xx} \overset{g_2}{\longmapsto} -\sigma_{\downarrow}^{y;xx}.
\label{eq:y;xx}
\end{equation}
This situation is shown in Fig.~\ref{fig:fig1}(a).
When $g_2$ is broken, for instance by strain, such constraints are lifted. The two magnetic sublattices are no longer connected by $g_2$, allowing $\Delta_\uparrow\neq\Delta_\downarrow$. 
To gain intuition on how the symmetry-breaking-induced SGA is tied to the behavior of the nonlinear photocurrents, consider $\Delta_a >0$, an incident light with a frequency $\Delta_\uparrow > \hbar \omega \gtrsim \Delta_\downarrow$. In this regime, the photocurrents are generated mainly in the spin-down bands, as shown in Fig.~\ref{fig:fig1}(b), generating a charge current as well as a spin current flowing along the same spatial direction. Note that the spin selectivity here arises solely from the energy window for (near-)resonant interband transitions. We assume that the light polarization itself does not selectively address one spin channel over the other, which is generally reasonable in systems with mixed orbital character, where strict polarization selection rules are absent. This is indeed the case for the example material discussed below. Thus, the $g_2$ symmetry-imposed cancellations are lifted, introducing both charge and spin photocurrent components that are previously forbidden. Reversing the sign of the SGA selects a frequency window in which the spin-up bands instead governs the photocurrent generation, enabling the same symmetry-breaking-induced charge and spin photocurrent components but with opposite signs. 


While the SGA provides a simple band-edge indicator of spin-group symmetry breaking and captures the nonlinear optical responses near the absorption threshold, a more precise description applicable over a broader frequency range is obtained by introducing the imbalance of the spin-resolved (optical) joint density of states (JDOS)~\cite{Han2011, Wong2025, Tanaka2024, Chen2025} $\Delta \rho (\omega)= \int_\mathbf{k} \Delta \rho (\omega, \mathbf{k}) =  \int_\mathbf{k} \left[\rho_\uparrow (\omega, \mathbf{k}) - \rho_\downarrow (\omega, \mathbf{k}) \right]$, where 
\begin{equation}
    \rho_\sigma(\omega, \mathbf{k}) = \sum_{n \in \text{VB}, m \in \text{CB}} 
\delta [ \omega^\sigma_{mn}(\mathbf{k}) - \omega]
\end{equation}
quantifies the phase space available in the spin-$\sigma$ channel for interband transitions at the given photon energy.
We have assumed temperatures much lower than the band gap, such that valence bands are fully occupied and conduction bands are empty.
From the general expressions for the nonlinear optical conductivity (see Appendix.~\ref{sec:NLORs}), it is evident that these response functions consist of the quantum geometry of the bands weighted by the available interband-transition phase space~\cite{sipe2000, Aversa1995,bpve1981}. Consequently, in each spin channel, JDOS directly determines the magnitude and sign of the photocurrents.
In AMIs with intact spin-group symmetry, $\Delta \rho = 0$. Upon (weak) spin-group symmetry breaking, a finite $\Delta \rho$ arises and transforms under symmetry operations in the same way as SGA. The trilinear coupling constructed for $\Delta_a$ can therefore be extended to $\Delta \rho$, which underlines shear strain as a knob to control the sign of the symmetry-breaking-induced spin and charge photocurrents.

\textit{Material system---}To demonstrate these concepts, we present the results of first-principle calculations of an anti-ferroelectric $d$-wave altermagnet, CuWP$_2$S$_6$, a 2D material recently proposed to host altermagnetism coupled to its ferroelectric properties~\cite{PhysRevLett.134.106801}.
Several compounds of the thiophosphate family~\cite{Susner2017} have been experimentally realized~\cite{Brec1986} and magnetically characterized~\cite{Wiedenmann1981,Joy1992,Wang2023,Yang2025}, suggesting \cwps~to be a promising material for further experiments. The computational details are in Appendix.~\ref{sec:comp_details}.

The magnetic unit cell of \cwps~is shown in Fig.~\ref{fig:fig2}(a-b).
The magnetic atoms, W, are sanwiched between two S planes, forming a planar structure with broken inversion symmetry.
The magnetic easy axis is along the $x$ direction, and the two magnetic sublattices are connected by a two-fold screw rotation. The collinear arrangement endows the system with the symmetry $g_1$, as defined in the previous section, while the two-fold screw rotation corresponds to the symmetry $g_2$, also defined previously, combined with a half unit cell lattice translation $\mathbf{t}$ along the $x$ direction.
The crystal belongs to the $P2_1$ space group, with the magnetic arrangement described by the spin space group $1.4.1.1$ in the absence of spin-orbit coupling~\cite{chen_unconventional_2025,PhysRevX.14.031038}, generated by the operations $g_1$ and $g_2 \textbf{t}$.
DFT calculations shown in Fig.~\ref{fig:fig2}(d) exhibit planar $d$-wave isoenergy surfaces~\cite{PhysRevX.12.031042}, which is prominent around $S$ at $E=E_f-0.2$~eV, and is also visible around $X$ with a smaller spin splitting.
The two symmetry generators in \cwps~impose that
\begin{equation}\label{eq.symmetry}
\begin{aligned}
\epsilon_\uparrow(k_x,k_y) &= \epsilon_{g_2\uparrow}(g_2k_x,g_2k_y)
= \epsilon_\downarrow(k_x,-k_y),\\
\epsilon_\uparrow(k_x,k_y) &= \epsilon_{g_1g_2\uparrow}(g_1g_2k_x,g_1g_2k_y)
= \epsilon_\downarrow(-k_x,k_y),
\end{aligned}
\end{equation}
which are consistent with the DFT band structure shown in Fig.~\ref{fig:fig2}(c): Band splitting occurs along the $S$--$\Gamma$--$S'$ path, while spin degeneracies are preserved along the $Y$--$\Gamma$--$X$--$S$ path. 

We compute all optical responses via projected Wannier functions. The band structure around the Fermi level is fitted well via $d$-orbitals centered on W sites~\cite{PhysRevB.107.235135}. The optical conductivities are computed in spin-up and spin-down channels separately according to equations ~\ref{eq:shift} and ~\ref{eq:injection}, and then added/subtracted to obtain charge/spin conductivities as shown in Fig.~\ref{fig:fig2}(f).  As discussed above, spin-group symmetries restrict collinear altermagnets to linear shift and circular injection channels, with spin and charge photocurrents generated by a given light polarization flowing along orthogonal directions. The nonzero components are shown in Fig.~\ref{fig:fig2}(f). 

Upon the application of a shear strain $\epsilon_{xy}$, 
the band structure 
exhibits spin-dependent shifts, as shown in Fig.~\ref{fig:fig2}(e).
The shifts reverse when the strain is applied with an opposite sign while maintaining the N\'eel order,
confirming the trilinear coupling between strain, N\'eel order, and SGA. The $d$-wave isoenergy surface in Fig.~\ref{fig:fig2}(d) also shows imbalanced spin-up and spin-down contours. 

\textit{NLOR under spin-group symmetry breaking---}
As illustrated in Fig.~\ref{fig:fig1}(b), additional spin and charge photocurrent components are activated by strain-induced spin-group symmetry breaking, with representative examples shown in Fig.~\ref{fig:fig2}(g-h).
These induced responses reverse sign upon sign changing of the shear strain $\epsilon_{xy}$, while preserving their magnitudes. 
The odd parity of both the spin and charge current under a sign flip of the strain is transparent from a symmetry perspective.  Under a moderate shear strain strength, the atomic positions relative to unit cell remain unchanged and lattices under tensile and compressive strains can be related by the spin-group symmetry $g_2$ (See Appendix.~\ref{sec:strain_mapping}). At a given optical frequency, the NLORs in these two strained states are symmetry related:
\begin{equation*}
\begin{split}
    & \sigma_\text{charge}^{y;xx}(\epsilon_{xy}) = \sigma_{\uparrow}^{y;xx}(\epsilon_{xy}) + \sigma_{\downarrow}^{y;xx}(\epsilon_{xy}) \\ 
    & \overset{g_2}{\longmapsto}  -\sigma_{\downarrow}^{y;xx}(-\epsilon_{xy}) - \sigma_{\uparrow}^{y;xx}(-\epsilon_{xy}) = -\sigma_\text{charge}^{y;xx}(-\epsilon_{xy}),\\
    &\sigma_\text{spin}^{x;xx}(\epsilon_{xy}) = \sigma_{\uparrow}^{x;xx}(\epsilon_{xy}) - \sigma_{\downarrow}^{x;xx}(\epsilon_{xy}) \\ 
    & \overset{g_2}{\longmapsto} \sigma_{\uparrow}^{x;xx}(-\epsilon_{xy}) -  \sigma_{\downarrow}^{x;xx}(-\epsilon_{xy}) = -\sigma_\text{spin}^{x;xx}(-\epsilon_{xy}).
\end{split}
\end{equation*}

\begin{figure}
    \centering
    \includegraphics[width=\linewidth]{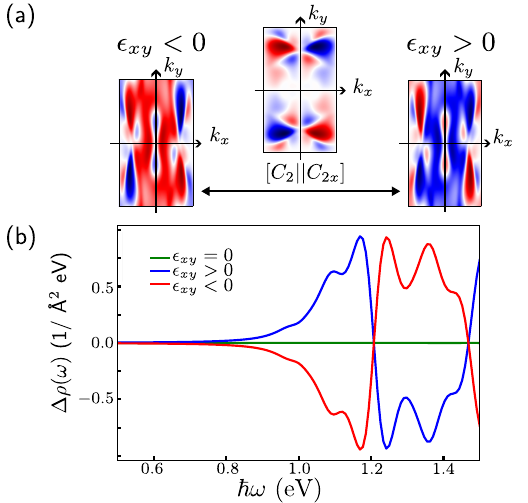}
    \caption{(a) Momentum-resolved imbalance in the spin-resolved joint density of states $\Delta \rho (\mathbf{k}, \omega)$ at fixed frequency $\hbar\omega = 1.25$ eV for pristine and strained unit cells. The unit cells under positive and negative shear strain are related by altermagnetic symmetries, which manifest in the corresponding $\Delta \rho$ patterns.
(b) Momentum-integrated $\Delta \rho (\omega)$ as a function of the incident photon energy, which is odd under strain reversal over the entire frequency range. }
    \label{fig:fig4}
\end{figure}

Microscopically, shear strain modifies the electronic structure and controls nonlinear optical responses through the induced imbalance of the spin-resolved JDOS $\Delta \rho$. 
In Fig.~\ref{fig:fig4}, we plot the momentum-resolved $\Delta \rho(\mathbf{k})$ and integrated $\Delta \rho$ in \cwps~monolayer at a given energy transfer. The momentum-resolved plot in the unstrained state exhibits the $d$-wave pattern respecting the altermagnetic symmetries $g_1$ and $g_2$. Under shear strain, $g_2$ connects the positively and negatively strained states via $\rho_\uparrow(\mathbf{k},\epsilon_{xy}) \overset{g_2}{\longmapsto} \rho_\downarrow(g_2\mathbf{k},-\epsilon_{xy})$, and yields
\begin{equation}\label{eq:JDOS_strain}
    \Delta \rho(\mathbf{k},\epsilon_{xy}) = \rho_\uparrow(\mathbf{k},\epsilon_{xy}) - \rho_\downarrow(\mathbf{k},\epsilon_{xy})=-\Delta \rho(g_2\mathbf{k},-\epsilon_{xy}),
\end{equation} 
as seen in Fig.~\ref{fig:fig4}(a).
Namely, $\Delta \rho (\mathbf{k})$ at $(k_x,k_y)$ under negative $\epsilon_{xy}$ and $(k_x,-k_y)$ under positive $\epsilon_{xy}$ have opposite signs. We assume that shear strain mainly shifts the bands, without significantly altering their quantum geometry.
The sign flip in $\Delta \rho (\mathbf{k})$ implies a switch in the spin channel that dominates the optical excitations. 
Along the direction where the photocurrents in the spin-up and spin-down channels are parallel to each other, the strain-induced spin imbalance generates a spin photocurrent with a sign odd under strain reversal. Meanwhile, along the direction where photocurrents in the spin-up and spin-down channels are antiparallel, the strain-induced spin imbalance generates a charge photocurrent that is likewise odd under strain. This microscopic picture is a direct extension of the SGA-based intuition presented earlier.
Integrating over the entire Brillouin zone results in an overall sign reversal in $\Delta \rho$ under positive and negative strain, as shown in Fig.~\ref{fig:fig4}(b). 
The strain and frequency dependence of $\Delta \rho$ is highly consistent with the behavior of NLORs presented in Fig.~\ref{fig:fig2}(g-h).



 


 
\textit{Discussion---}Our results establish a general symmetry-based principle for controlling nonlinear optical responses in altermagnetic insulators (AMIs): spin-group–symmetry breaking gives rise to nonlinear charge and spin photocurrent channels with their sign deterministically locked to the applied perturbation. In particular, we show that shear strain enables such symmetry control through a trilinear coupling to the Néel order and the spin-resolved electronic asymmetry. The photocurrent behavior has an intuitive interpretation at the band edge in terms of spin-gap asymmetry, and more generally are governed by the spin-resolved imbalance of the optical joint density of states.
In the example of \cwps, although the strain-induced charge and spin photocurrents are weaker than the originally symmetry-allowed ones, their experimental detection can be enhanced by a polar-angle map through a rotational scan of the electric-field polarization (see Appendix.~\ref{sec:polar_angle}). By the same symmetry-based reasoning, spin-group symmetry breaking can also activate additional components of the second-harmonic-generation susceptibility and alter the corresponding polarization selection rules.

These results establish nonlinear optical responses as an informative probe of AMIs. Further, AMIs can be explored as a versatile platform for controllable spin and charge photogalvanic phenomena, opening new opportunities for optospintronic functionalities that do not rely on net magnetization or spin–orbit coupling, but are guided by spin-group symmetry. Our analysis can be straightforwardly generalized to other symmetry-breaking perturbations, their specific form of coupling to the system and resulting responses are dictated by their symmetry transformation properties.
More broadly, our work identifies a general route for tuning physical responses governed by spin-resolved electronic asymmetry. Such asymmetry can be induced by a wide class of symmetry-breaking effects. Especially near the onset of interaction-driven spontaneous symmetry breaking, enhanced susceptibility is expected to strongly amplify the effectiveness of such perturbative control. 



\textit{Acknowledgements---}We thank U. Nitzsche for technical assistance, Liuyan Zhao, Jiangxu Li, Anshul Kogar, Yang Zhang, and Libor Šmejkal for fruitful discussions. This work is supported by the Leibniz Association through Leibniz Competition Project No. J200/2024.

\textit{Data availability---}The data are available from the authors upon reasonable request.

\bibliography{references} 

\vspace{10pt}

\appendix
\setcounter{secnumdepth}{2}

\renewcommand{\thesection}{\Alph{section}}      
\renewcommand{\thesubsection}{\thesection.\arabic{subsection}}  

\section{Definition of NLORs}
\label{sec:NLORs}
Equation~\ref{eq:photocurrent_current} in the main text gives the current response to the second order in the incident electric field. As mentioned there, the DC response can be separated into the shift and injection currents, which can be computed in terms of band geometric quantities as follows
\begin{subequations}\label{eq:response}
\begin{align}
\sigma_{\mathrm{shi}}^{c;ab}(\omega) &= -\frac{\pi e^3}{\hbar^2} 
\int_\mathbf{k} \sum_{n,m} f_{nm} 
\Bigl(R_{mn}^{c,a} - R_{nm}^{c,b}\Bigr) \nonumber \\
& \quad \times r_{nm}^b r_{mn}^a \, \delta(\omega_{mn} - \omega),
\label{eq:shift} \\
\sigma_{\mathrm{inj}}^{c;ab}(\omega) &= -\tau \frac{2\pi e^3}{\hbar^2} 
\int_\mathbf{k} \sum_{n,m} f_{nm} \, \Delta_{mn}^c  \nonumber \\
& \quad \times r_{nm}^b r_{mn}^a \delta(\omega_{mn} - \omega) \label{eq:injection}.
\end{align}
\end{subequations}
Here, the superscripts $a,b,c$ denote Cartesian components, and $\sigma^{c;ab}$ is the second-order conductivity tensor along direction $c$ associated to the components $a$ and $b$ of the electric field. We define $\int_{\mathbf{k}} = \int d^d k/(2\pi)^d$, $f_{nm} = f_n - f_m$ as the difference of Fermi–Dirac occupations, and $r_{mn}^a = \langle m | i \partial_a | n \rangle$ as the interband Berry connection. The energy difference between bands is $\hbar \omega_{mn} = \hbar(\omega_m - \omega_n)$, and $\tau$ denotes the momentum relaxation time.

The shift vector is defined as $R_{mn}^{c,a} = r_{mm}^c - r_{nn}^c + i \partial_c \log r_{mn}^a$, while $\Delta_{mn}^c = v_{mm}^c - v_{nn}^c$ is the difference in band velocities along the $c$ direction. All integrand quantities are implicitly momentum dependent, and this dependence is suppressed for notational simplicity.

\section{Computational details}
\label{sec:comp_details}
We perform scalar-relativistic DFT calculations for the CWPS monolayer using the Full-Potential Local-Orbital (FPLO) code, version 22.02~\cite{PhysRevB.59.1743}. The Perdew-Burke-Ernzerhof implementation of the generalized gradient approximation (GGA) is employed. A $k$-mesh of $6 \times 9 \times 1$ is used for numerical integration over the Brillouin zone. To account for the strongly correlated nature of the W-$5d$ orbitals, we include a GGA+$U$ correction with $U = 1$~eV~\cite{PhysRevLett.134.106801}. The local magnetic moment at each W site is approximately $2.5~\mu_B$.

\section{Mapping of strained unit cells}
\label{sec:strain_mapping}

Denote the strained lattice vectors by
\begin{equation}
\mathbf{a}_1(\epsilon_{xy}) = a (1, \epsilon_{xy}), \quad 
\mathbf{a}_2(\epsilon_{xy}) = b (\epsilon_{xy}, 1),
\end{equation}
and the atomic positions in fractional coordinates of the unit cell are
\begin{equation}
\mathbf{r}_1 = \frac{\mathbf{a}_2}{4}, \quad 
\mathbf{r}_2 = \frac{\mathbf{a}_1}{2} + \frac{3 \mathbf{a}_2}{4}.
\end{equation}
We consider the spin-group operation $g_2 = [C_2 \Vert C_{2x}\,\mathbf{t}]$, with
$\mathbf{t} = \mathbf{a}_1(-\epsilon_{xy})/2$, where $C_{2x}$ acts as
$(x,y)\rightarrow(x,-y)$. Applying $g_2$ exchanges the two sublattices while mapping
$\epsilon_{xy} \rightarrow -\epsilon_{xy}$.
 Acting on $\mathbf{r}_1$, we obtain
\begin{align}
g_2 \mathbf{r}_1(\epsilon_{xy}) &= C_{2x} \mathbf{r}_1(\epsilon_{xy})
+ \frac{\mathbf{a}_1(-\epsilon_{xy})}{2} \nonumber\\
&= \frac{b}{4}(\epsilon_{xy},-1)
+ \frac{\mathbf{a}_1(-\epsilon_{xy})}{2} \nonumber\\
&= \frac{\mathbf{a}_1(-\epsilon_{xy})}{2}
- \frac{\mathbf{a}_2(-\epsilon_{xy})}{4} \nonumber\\
&= \mathbf{r}_2(-\epsilon_{xy}),
\end{align}
where we have added a lattice vector to kg the position back into the unit
cell.

Similarly, for $\mathbf{r}_2$,
\begin{align}
g_2 \mathbf{r}_2(\epsilon_{xy}) &= C_{2x} \mathbf{r}_2(\epsilon_{xy})
+ \frac{\mathbf{a}_1(-\epsilon_{xy})}{2} \nonumber\\
&= \frac{a}{2}(1,-\epsilon_{xy})
+ \frac{3b}{4}(\epsilon_{xy},-1)
+ \frac{\mathbf{a}_1(-\epsilon_{xy})}{2} \nonumber\\
&\equiv \mathbf{r}_1(-\epsilon_{xy}),
\end{align}
again up to a lattice translation.

This explicitly demonstrates that the operation $g_2$ exchanges the two
sublattices between the tensile and compressive unit cells, thereby justifying
the relations used in the main text for the strain-dependent conductivities.
Although the choice of the translation vector $\mathbf{t}$ may appear arbitrary,
note that in the zero-strain limit $g_2$ reduces to the altermagnetic symmetry.

\label{sec:polar_angle}
\begin{figure}[ht]
    \centering
    \includegraphics[width=\linewidth]{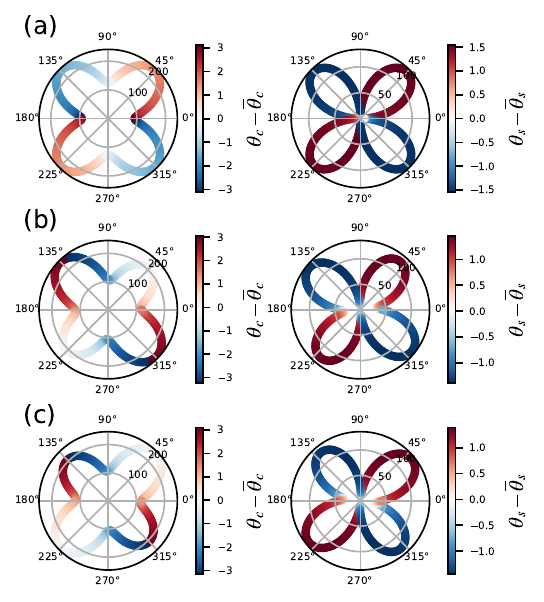}
    \caption{Polar angle plots of the charge  (left panels) and spin (right panels) currents generated with linearly polarized light at $\hbar\omega = 1.2$ eV for: (a) $\epsilon_{xy} = 0$, (b) $\epsilon_{xy} > 0$ and (c) $\epsilon_{xy} < 0$. The color of the points indicate the value of the angle associated to the direction $\theta$ of the generated current, with respect to the its average value.}
    \label{fig:fig3}
\end{figure}
In Fig.~\ref{fig:fig3}(a--c) we show polar plots of the charge current $J_c(\phi)$, left panels, and spin current $J_s(\phi)$, right panels, generated by linearly polarized light with electric field $\mathbf{E}=E(\cos\phi,\sin\phi,0)$ at a fixed frequency $\hbar\omega = 1.2$~eV. The data points are colored according to the shifted current directions $\theta_c-\overline{\theta}_c$ and $\theta_s-\overline{\theta}_s$, which share the same color scale, where $\tan\theta_c = J_c^y/J_c^x$, $\tan\theta_s = J_s^y/J_s^x$, and $\overline{\theta}_c$ and $\overline{\theta}_s$ denote the angular averages over $\phi$.

\section{Polar angle plot}
In Fig.~\ref{fig:fig3}(a), the unstrained system exhibits a perfectly
symmetric $d$-wave--like pattern. In contrast, the pattern is distorted and rotated in
Fig.~\ref{fig:fig3}(b-c), with the
rotation occurring in opposite directions for oppositely applied strains.

\newpage

\end{document}